# A GEM-based detector for detection and imaging sparks and flames

G. Volpe[1] and V. Peskov[2]

[1]Dipartimento Interateneo di Fisica \M. Merlin" and Sezione INFN, Bari, Italy
[2] CERN and  Institute for Chemical Physic Russian Academy of Sciences

## Abstract

Earlier we have developed and successfully tested a RICH detector prototype consisting in a CsI coated triple GEM operated in gas flushed mode

In the given work, a modified version of this detector for a completely different application - fire safety- is presented.  The detector operates in sealed mode and is combined with an optical system and a narrow band filter

As a photosensitive element, a CsI photocathode coated with a thin layer of ethylferrocene was used. This detector is almost 1000 times more sensitive than the best commercial flame sensor; it has 100 times better time resolution and allows determining the location where the spark or flame appears

## I.   Introduction

It is very important to determine a fire hazard on its early stage. There are several commercial sensors capable to detect appearing of small flames. The most sensitive among them are those who operate in UV region of spectra: 185-250 nm. In this wavelength interval, all the flames in air emit quite strongly, whereas the Sunlight is blocked by the ozone in the upper layer of the atmosphere. An example could be Hamamatsu R2868 UVtron, which is in fact a gaseous detector with a metallic photocathode. Since this detector operates in digital mode, it cannot distinguish between a single photon, a cosmic ray or a spark.

Our ideas were to replace metallic with a CsI photocathode, which is on orders of magnitude more sensitive than the metallic one, and   to use a GEM detector, which has imaging capability.

To materialise these ideas we used a detector, which we develop earlier for the RICH applications. It consisted of a CsI- coated triple GEM operated in gas-flushed mode [1].

## II.   Position-sensitive detector for flame imaging

The imaging flame detector is a sealed gas chamber with a UV transparent window (see Figures.1and 2).  Inside the chamber a triple resistive GEM detector is installed, whose top GEM is coated with a CsI layer 0.4 μm thick (see Figures 1and 2). This GEMs has resistive electrodes instead of metallic one, making it to be spark-protected [2]. Each resistive GEM has a 10x 10 cm² active area, 0.45 mm thickness, 0.4 mm hole diameter and 0.8 mm pitch (Figure 3). Below the GEM, a pad readout plane is located (pads size 8x8 mm²). The gas chamber was filled either with Ne+10%CH$_4$ or with Ne+10%CF$_4$ at 1 atm pressure.

In this device, a UV photon can extract an electron from the CsI photocathode that is deposited on top of the first GEM upper surface. The electron is led by the electric field action to the nearest hole, where it experiences the first amplification; then the avalanche electrons undergo a second amplification in the following GEM (and more, depending on the number of GEM foils) and they finally induce a signal on the pad-type readout plate.

For imaging purpose, in the front of the entrance widow a lens is placed, such that the top GEM surface, which is coated with the CsI layer, is in its focal plane. In this arrangement any objects, which are located at a distance much larger than the lens focus, are projected on the CsI surface.

Combined with a proper electronics, this detector allows visualization of flames or sparks

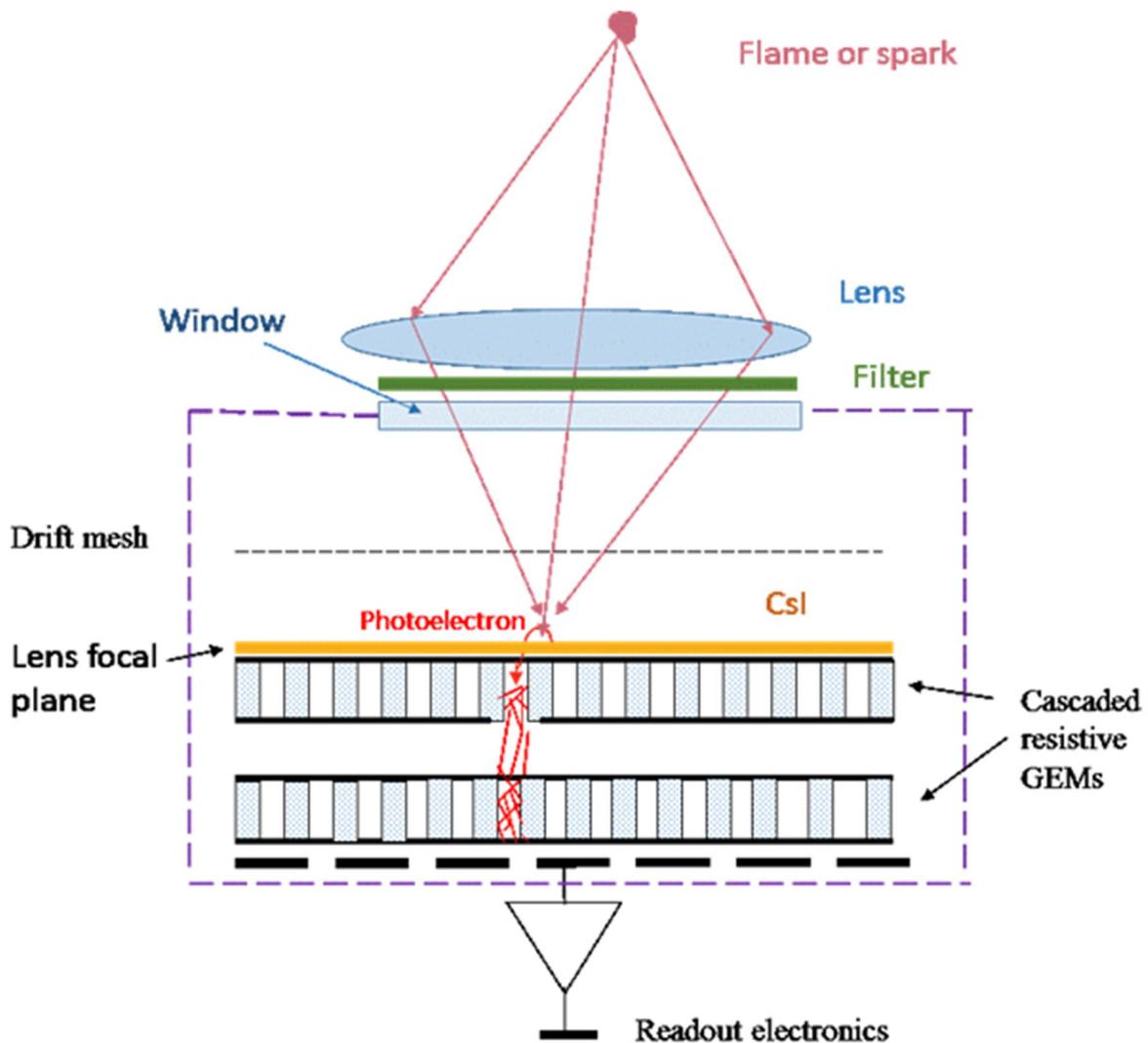

Figure 1. Working principle of a cascaded resistive GEM combined with a CsI photocathode; the optical system is also shown

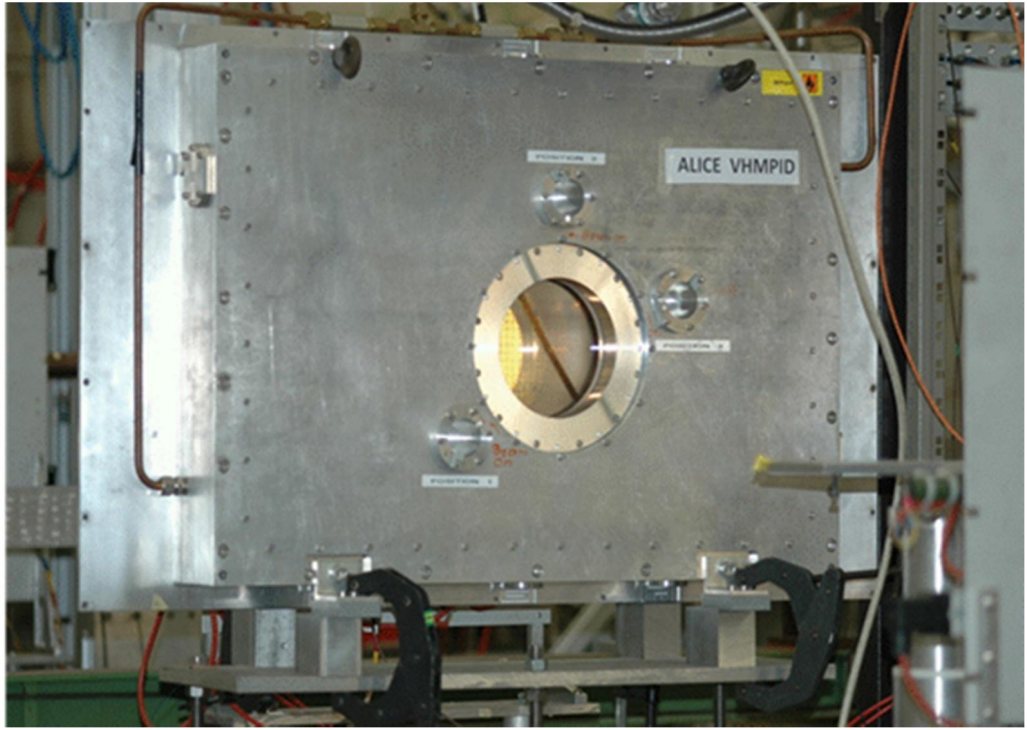
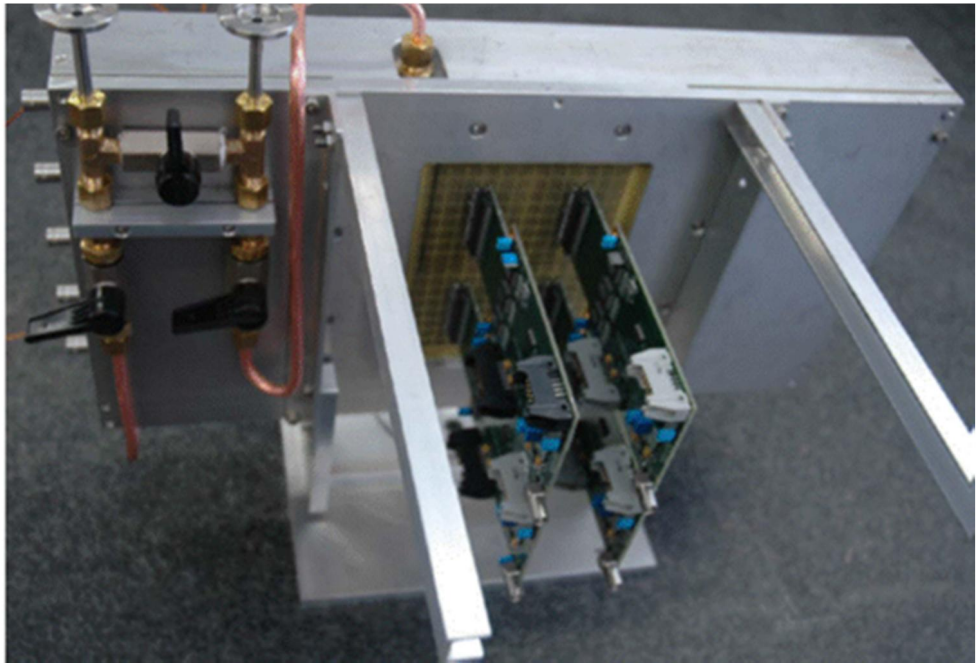

Figure 2. a) Photographs of the front view of a flame detector, employing triple resistive GEM, with the top electrode coated with a CsI layer. A $CaF_2$ window in the centre of the front flange, facing the particle beam; can be clearly seen, as well as three windows for radioactive sources used for preliminary tests b) Back view of the detector, showing the front-end electronics connected to the readout pad plane

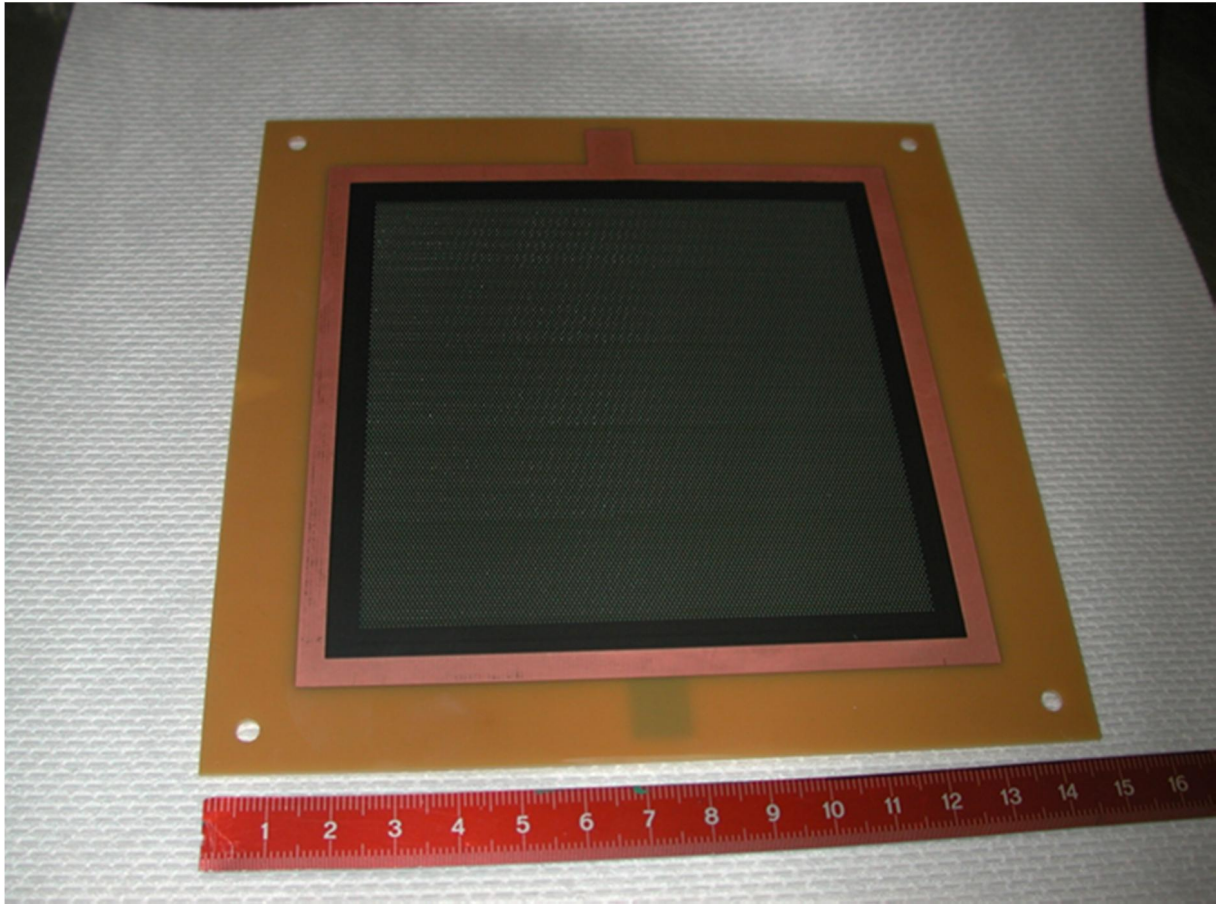

Figure 3. Photograph of a spark-protected resistive thick GEM

During image taking, the triple resistive GEM operated with a reversed drift electric field (around 200 V), to enhance the photoelectron extraction efficiency from the CsI cathode [3] and to supress undesirable signals from cosmic or natural radioactivity at an overall gain of ~$10^5$

To adopt this detect for the fire safety applications several important modifications were done:

1) A narrow -band filter was placed in front of the input window
2) To compensate the sensitivity loss due to the filter, a CsI photocathode was coated with a thin ethylferrocene layer which enhanced its quantum efficiency in the interval 190-220 nm [4]
3) It operated in only in proportional mode allowing to detect sparks if analog signals are used or obtain digital image of flames

Thanks to these modifications, this detector is 500 times more sensitive than the best commercial flame sensors, it has much faster respond (time resolution few µs), it allows to determine the direction, where a spark or a flame appears (Figure 1) and it is able to operate even in direct sunlight illumination. Moreover, two such detectors located in different places and operating simultaneously can accurately determine flame position in 3D
Provided with an appropriate algorithm for pattern recognition, this detector could achieve a high rejection of false signals, making this flame-detection system very robust.

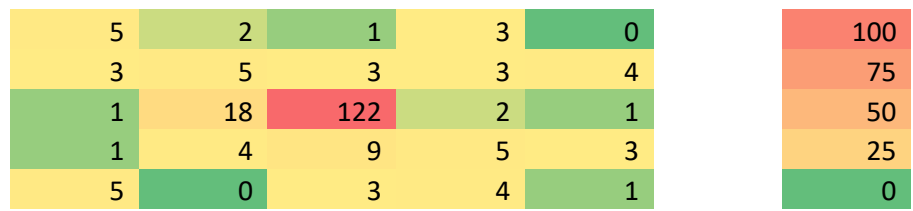

a)

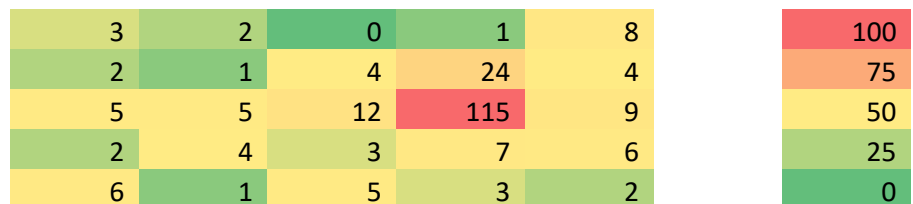

b)

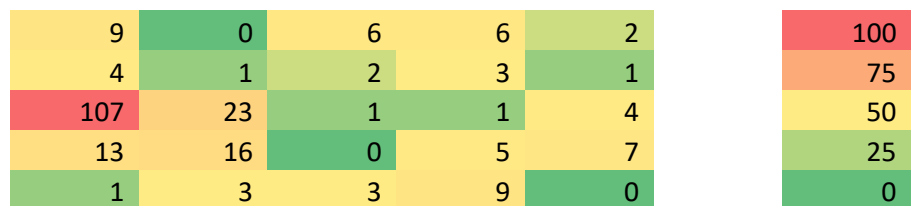

c)

Figure 4. Left: digital images (or number of counts per readout pad measured during 10 s) of a candle flame placed 15 m away from the detector in three different positions:

   a) On the line perpendicular to the window surface and passing through its centre
   b) Shifted 1.3m to the right
   c) Moved to 2.5 m to the left

Right: the colour scale is shown

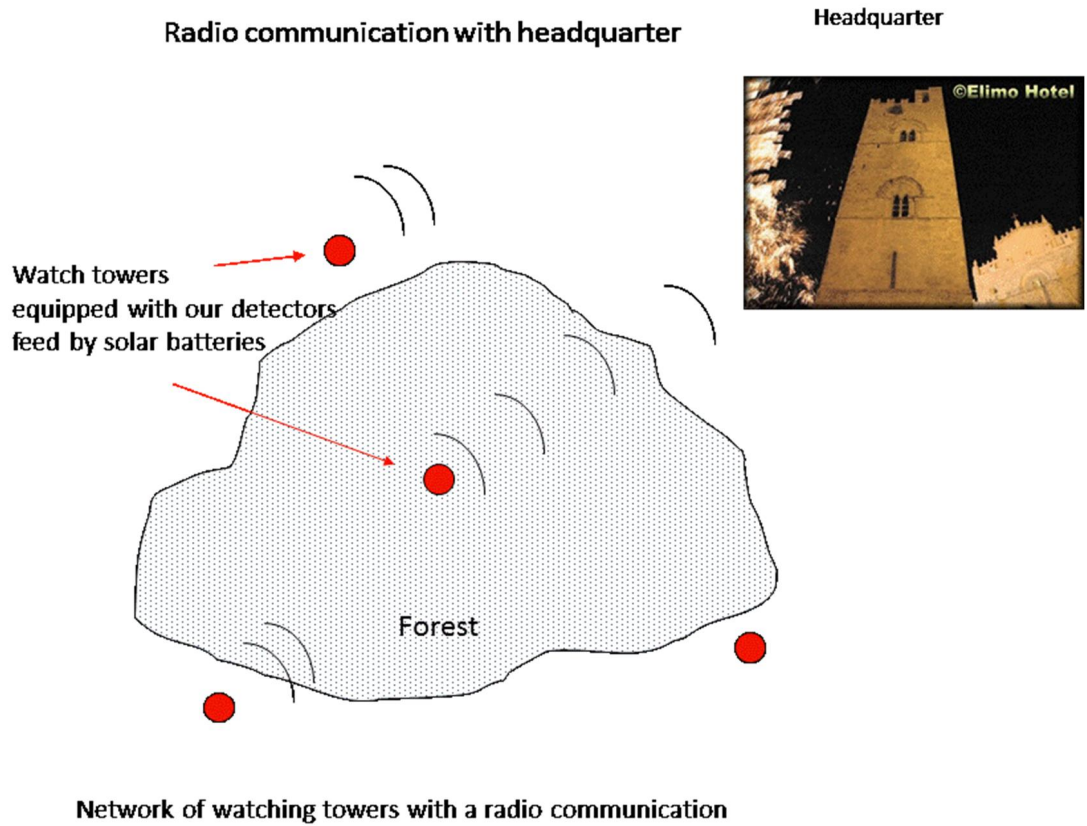

Figure 5. Possible arrangement of flame imaging detectors allowing to determine the flame position

### III. Conclusions

Preliminary tests of the developed prototype demonstrated that GEM-based UV imaging detectors could be useful for the for the fire safety monitoring of forests, large holls, industrial installations having fire risks etc.

**Acknowledge**:

This work was partially supported by: Fondo di Sviluppo e Coesione 2007-2013 - APQ Ricerca Regione Puglia "Programma regionale a sostegno della specializzazione intelligente e della sostenibilita sociale ed ambientale - FutureInResearch"